\PassOptionsToPackage{bookmarks=false}{hyperref}
\documentclass[sigconf,9pt,screen]{acmart}
\usepackage{graphicx,amsmath,array,booktabs}
\usepackage{balance}
\usepackage{algorithm}
\usepackage{cleveref}
\AtBeginDocument{\hypersetup{linkcolor=blue,citecolor=blue,urlcolor=blue,filecolor=blue}}

\newcommand{\minisection}[1]{\vspace{.06in}\noindent{\textbf{#1}}.}

\copyrightyear{2027}
\acmYear{2027}
\setcopyright{cc}
\setcctype{by}
\acmConference[ASPDAC '27]{32nd Asia and South Pacific Design Automation Conference}{January 25--28, 2027}{Tokyo, Japan}
\acmBooktitle{32nd Asia and South Pacific Design Automation Conference (ASPDAC '27), January 25--28, 2027, Tokyo, Japan}
\acmDOI{10.1145/3840404.3860937}
\acmISBN{979-8-4007-2923-2/2027/01}
\begin{document}
\title[CU-3D: A Scalable Legality-Aware RTL-to-GDS Flow]{CU-3D: A Scalable Legality-Aware RTL-to-GDS Flow for 3D-IC Physical Design}

\author{\texorpdfstring{%
    \large\normalfont
    Mingjun Wang\textsuperscript{1,*},\quad
    Guanqi Li\textsuperscript{2,5,*},\quad
    Yihan Wen\textsuperscript{1},\quad
    Lancheng Zou\textsuperscript{1},\quad
    Mingjun Li\textsuperscript{1},\quad
    Feng Gu\textsuperscript{1}\\
    Fengrui Liu\textsuperscript{2,5},\quad
    Jianan Mu\textsuperscript{2},\quad
    Yuxuan Zhao\textsuperscript{3},\quad
    Tsung-Yi Ho\textsuperscript{1},\quad
    Huawei Li\textsuperscript{2,5},\quad
    Sung Kyu Lim\textsuperscript{4},\quad
    Bei Yu\textsuperscript{1}\\[3pt]
    {\normalsize\normalfont
    \textsuperscript{1}The Chinese University of Hong Kong\quad
    \textsuperscript{2}Institute of Computing Technology, Chinese Academy of Sciences, Beijing, China}\\
    {\normalsize\normalfont
    \textsuperscript{3}The Chinese University of Hong Kong, Shenzhen\quad
    \textsuperscript{4}University of Southern California}\\
    {\normalsize\normalfont
    \textsuperscript{5}University of Chinese Academy of Sciences, Beijing, China}%
}{Mingjun Wang, Guanqi Li, Yihan Wen, Lancheng Zou, Mingjun Li, Feng Gu, Fengrui Liu, Jianan Mu, Yuxuan Zhao, Tsung-Yi Ho, Huawei Li, Sung Kyu Lim, Bei Yu}}
\thanks{\textsuperscript{*}Mingjun Wang and Guanqi Li contributed equally to this work.}
\renewcommand{\shortauthors}{Wang et al.}
\date{}
\begin{abstract}
    In face-to-face (F2F) 3D ICs, data, clock, and power connections compete for hybrid-bond (HB) sites, while clock buffers require access to a connected power grid (PG).
    We present CU-3D, a legality-aware RTL-to-GDS physical design flow built on commercial EDA tools.
    Using a 17-metal ASAP7-based technology model, CU-3D screens complete HB footprints against the constructed power grid on both dies.
    It supports 2D placement reuse and top-die-first re-placement. In re-placement, top-die pins and clock loads guide provisional joint data and clock HB assignment and subsequent bottom-die placement.
    DATA HBs can be reassigned to accommodate clock demand while reserving capacity for PG HBs.
    Clock-tree synthesis and routing preserve the selected HB sites and stacks, and buffer insertion checks supply access and connectivity.
    The post-route layouts for ten designs, including a 133-macro processor, have zero backend design-rule violations and pass signal-connectivity checks.
    Across five 2D/3D comparisons, CU-3D improves worst setup slack by 1--28 ps, reduces per-die footprint by 35.6--40.0\%, and lowers tool-reported power by 4.9--15.3\%.
    On JPEG, joint HB assignment improves worst setup slack by 87 ps over data-only assignment, and tier-wide partitioning area bounds yield a further 346 ps gain.
\end{abstract}
\maketitle
\pagestyle{plain}
\thispagestyle{plain}
\section{Introduction}
\label{sec:intro}

Face-to-face (F2F) hybrid bonding connects dies through their interconnect faces~\cite{soic2019}. Submicron hybrid bonds (HBs) can replace long planar wires with short vertical connections~\cite{imec2023hybrid}. The timing benefit depends on the wires from cell pins to the HB sites on both dies and on the clock paths to the launch and capture registers. Evaluating these effects requires a powered, clocked, and routed layout.

Prior work has developed methods for reusing 2D placement and routing engines~\cite{shrunk2d,compact2d,snap3d}, assigning nets to bonding sites~\cite{btassign,anchorgrid}, synthesizing cross-die clock trees~\cite{flexicts}, and building power delivery networks~\cite{dpo3d}. These operations share physical resources. The constructed power grid (PG) can obstruct a DATA or CLOCK HB stack, and a DATA HB can occupy a site needed by a clock sink group. During clock-tree synthesis (CTS), a buffer placed in a legal row still needs continuous paths from both supply pins to the PG.

\begin{figure}[tb!]
    \centering
    \includegraphics[width=.918\columnwidth]{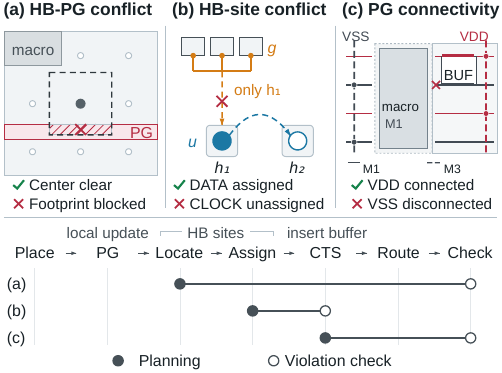}
    \caption{
        Physical constraints that link early planning to downstream checks.
        (a) HB-PG conflict: PG overlaps the footprint, not its center.
        (b) HB-site conflict: DATA $u$ occupies CLOCK group $g$'s only site $h_1$; $u$ can move to $h_2$. Rectangles/dots: D flip-flops (DFFs)/clock pins.
        (c) PG connectivity: a macro interrupts M1 rails (solid); BUF's VSS rail is disconnected from M3 (dashed). Dots: layer connections. Filled/open timeline markers: planning/violation checks.
    }
    \Description{Three schematic plan views show an HB footprint conflict, an HB-site assignment conflict, and disconnected PG rails. In c, a macro and its M1 clearance interrupt four alternating supply rails. The buffer is outside the macro, fills one legal row, and contacts both local rails. Dashed M3 straps connect to the PG network outside the window. Dots denote layer connections, not individual vias. The right VDD rail connects to its strap, but the right VSS fragment has no path to the VSS strap. Macro-internal PG is omitted. In the timeline, Locate denotes initial HB-site selection and Assign maps requests to sites.}
    \label{fig:chain}
\end{figure}

Figure~\ref{fig:chain} links planning to implementation: complete HB footprints must clear both dies' obstructions, DATA and CLOCK HBs need distinct sites, and inserted buffers need connected VDD and VSS supplies. In (c), legal buffer placement does not ensure a ground connection.

\begin{table*}[t]
\caption{What representative 3D methods physically build and check.}
\label{tab:scope}
\centering
\begingroup
\footnotesize
\setlength{\tabcolsep}{3pt}
\renewcommand{\arraystretch}{1.08}
\begin{tabular*}{\textwidth}{@{\extracolsep{\fill}}llllll@{}}
\toprule
Method & Interface representation & Clock construction & PG & Macros & Reported DRC \\
\midrule
Shrunk-2D~\cite{shrunk2d} & Monolithic 3D / MIVs & 2D CTS, tier split & -- & Yes & -- \\
Compact-2D~\cite{compact2d} & Combined metal stack & 2D CTS, tier split & -- & Yes & -- \\
Snap-3D~\cite{snap3d} & Tier rows / F2F vias & Combined CTS & -- & Yes & -- \\
Macro-3D~\cite{macro3d} & Macro-on-logic & Measured clock tree & -- & Yes & -- \\
Hetero-3D~\cite{hetero3d} & Mixed-node F2F & CTS & Built & Yes & -- \\
BTAssign~\cite{btassign} & Legal HB mesh & -- & -- & -- & -- \\
FlexiCTS~\cite{flexicts} & HBT allocation & CPPR-aware 3D CTS & -- & -- & -- \\
AnchorGrid~\cite{anchorgrid} & Pad array & Root-balanced & -- & -- & 224 (CPU2) \\
Open3DBench~\cite{open3dbench} & Vias / virtual HBs & MoL: CTS; LoL: ideal & -- & Yes & not clean (LoL) \\
DPO-3D~\cite{dpo3d} & F2F power network & -- & Built & -- & not clean \\
\midrule
\textbf{CU-3D} & \textbf{HB + local-via stacks} & \textbf{TOP then BOTTOM} & \textbf{Built} & \textbf{0--133} & \textbf{0 (10 layouts)} \\
\bottomrule
\end{tabular*}
\par\smallskip\raggedright
PG: constructed power grid; --: unspecified scope. MoL/LoL: memory-/logic-on-logic. DRC counts use each work's backend checks.
\par\endgroup
\end{table*}

We present CU-3D, a legality-aware RTL-to-GDS physical design flow built on commercial EDA tools.
CU-3D derives HB-site capacity from the constructed PG and uses placement information to allocate that capacity before CTS and routing.
Its contributions are:
\begingroup
\setlength{\leftmargini}{5.08mm}
\begin{itemize}
\setlength{\itemsep}{0pt}
\setlength{\parsep}{0pt}
\setlength{\parskip}{0pt}
\item \textbf{HB-site screening.} CU-3D checks complete HB footprints on both dies and uses same-supply metal on both dies to identify candidate PG HB sites.
\item \textbf{Partitioning and placement.} Local area bounds support 2D placement reuse, while tier-wide bounds support top-die-first re-placement.
    In re-placement, top-die pin locations and clock loads guide provisional HB assignment and bottom-die placement.
\item \textbf{Joint DATA/CLOCK HB assignment.} CU-3D can reassign DATA HBs to accommodate CLOCK HBs while reserving capacity for subsequent PG HB assignment.
\item \textbf{CTS, PG, and routing.} CU-3D synthesizes the top-die clock subtrees before the bottom-die tree, checks supply access and connectivity for inserted cells,
    and preserves the selected HB sites and stacks through routing.
\end{itemize}
\endgroup

All ten post-route layouts, including a 133-macro processor, have zero backend design-rule violations and pass signal-connectivity checks.
\section{Related Work}
\label{sec:related}

\minisection{Reuse of 2D design tools}
Shrunk-2D~\cite{shrunk2d} scales cell and interconnect geometries to obtain a 2D placement for tier partitioning. Compact-2D~\cite{compact2d} scales interconnect RC parasitics, contracts the resulting layout to the F2F footprint, and optimizes the design after partitioning. Snap-3D~\cite{snap3d} contracts cell heights and encodes tiers as placement rows so that a commercial 2D placer can place cells across tiers. Cascade2D~\cite{cascade2d} and Pin-3D~\cite{pin3d} address monolithic 3D physical synthesis, with Pin-3D supporting timing optimization across dies. Macro-3D~\cite{macro3d} targets heterogeneous macro-on-logic F2F designs.

\minisection{HB assignment, clocking, and power delivery}
BTAssign~\cite{btassign} performs routing-aware assignment to legal HB sites. AnchorGrid~\cite{anchorgrid} represents cross-die nets with paired anchor cells that can move during pre-route timing optimization and are later assigned to a bonding grid. Kim and Park~\cite{kim2025unified} address fine-pitch HB assignment in a unified flow for homogeneous and heterogeneous integration. Jiang et al.~\cite{jiang2026partition} account for HB terminal demand during partitioning. FlexiCTS~\cite{flexicts} coordinates clock buffer placement and HB allocation, while DPO-3D~\cite{dpo3d} and Zhu et al.~\cite{zhu2022power} address power delivery. Related timing methods target early-stage timing prediction~\cite{distill3d}, timing-driven F2F placement~\cite{timing3dplacement}, and concurrent clock--data optimization~\cite{diffccd}.

\minisection{Implementation flows and benchmarks}
Hetero-3D~\cite{hetero3d} integrates mixed-node floorplanning and power delivery. Patnaik et al.~\cite{patnaik2018} report DRC-clean F2F layouts that use redistribution layers. Open3DFlow~\cite{open3dflow,open3dflow_artifact} provides an open-source 3D design platform, and Open3DBench~\cite{open3dbench} supports evaluation from partitioning and placement through routing, RC extraction, and thermal analysis. RosettaStone~\cite{rosettastone} addresses reproducibility in physical design benchmarking.

Table~\ref{tab:scope} summarizes the structures implemented and checks reported by representative methods. CU-3D derives legal HB sites from the constructed power grid, assigns DATA and CLOCK HBs jointly while reserving capacity for PG HBs, and checks VDD/VSS access and connectivity when inserting cells. The assembled post-route layout is then checked for design-rule violations and connectivity with the routed clock trees, PG, and HB stacks in place.
\section{Flow Overview}
\label{sec:overview}

\begin{figure*}[t]
\centering
\includegraphics[width=0.99\textwidth]{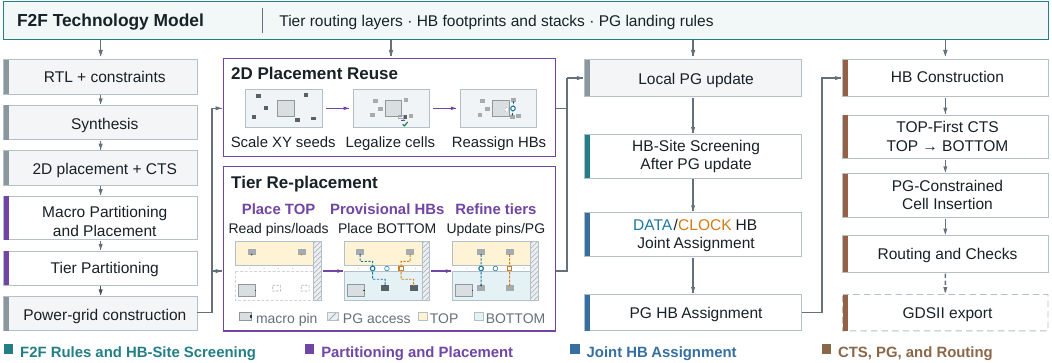}
\caption{CU-3D flow. Colors identify method chapters; gray marks shared steps. Provisional HBs use initial candidate sites; final assignment follows PG updates and HB-site re-screening. Dashed GDSII export requires complete stream and cell views.}
\Description{The original four-column flow is retained. Four bottom color keys reproduce the method chapter titles: F2F Rules and HB-Site Screening in teal, Partitioning and Placement in purple, Joint HB Assignment in blue, and CTS, PG, and Routing in brown. Gray marks RTL input, synthesis, reference 2D placement and CTS, power-grid construction, and local PG update. Macro and tier partitioning and both placement branches are purple. Provisional HBs guide BOTTOM placement using initial candidate sites. The technology band and HB-site screening are teal; screening is repeated after the local PG update. Joint DATA/CLOCK HB assignment and subsequent PG HB assignment form a contiguous blue pair. The rightmost construction, CTS, insertion, routing, and conditional export steps remain brown. Neutral arrows retain the original execution and data dependencies.}
\label{fig:overview}
\end{figure*}

Figure~\ref{fig:overview} outlines CU-3D. Reference 2D placement and CTS measure timing and cell-area overhead.

\textbf{F2F rules and HB-site screening} check complete HB footprints against both dies' geometry.

\textbf{Partitioning and placement} assign macros and cells to tiers. 2D placement reuse legalizes scaled reference coordinates; tier re-placement uses TOP pins and loads to guide provisional HBs and BOTTOM placement. Both update pins and PG before final HB-site screening.

\textbf{Joint HB assignment} selects DATA/CLOCK HBs jointly, then PG HBs at remaining legal sites. DATA includes control/I/O; PG connects VDD/VSS across dies. \textbf{CTS, PG, and routing} construct HBs, perform TOP-first CTS and PG-constrained cell insertion, then route and check the assembled layout. Post-route results include backend DRC, signal/PG connectivity, worst negative slack (WNS), and total negative slack (TNS).

\section{F2F Rules and HB-Site Screening}
\label{sec:model}

\begin{figure}[t]
\centering
\includegraphics[width=0.97\columnwidth]{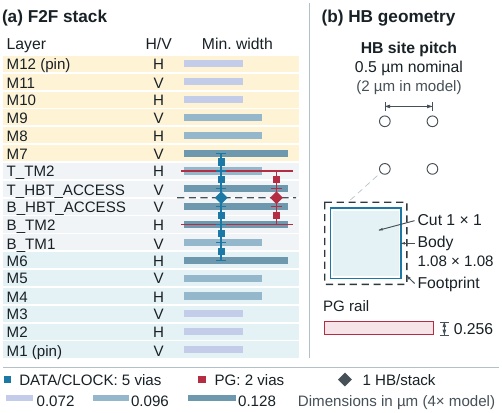}
\caption{F2F technology model. (a) F2F stack: layer order, preferred horizontal/vertical (H/V) routing, and minimum widths. DATA/CLOCK and PG use five and two local vias (squares), respectively, plus one HB (diamond) each. (b) HB geometry: XY cut, body, and clearance footprint. Dimensions use the $4\times$ model except nominal pitch; vertical spacing is schematic.}
\Description{A seventeen-metal layer-order view shows minimum widths and two conductor stacks, not physical thicknesses. DATA/CLOCK uses five local vias and one HB; PG uses two local vias and one HB. Squares denote local vias, diamonds the single bond in each stack. A separate XY plan distinguishes the bond cut, body, pitch, and full clearance footprint.}
\label{fig:pdk}
\end{figure}

\subsection{F2F Technology Model}
CU-3D uses an ASAP7-based technology model to screen candidate HB sites for spacing, enclosure, and access~\cite{asap7,asap7dtco}. The dies share an XY coordinate system and database, with separate routing layers and explicit cross-die nets in the timing graph. Figure~\ref{fig:pdk}(a) shows twelve intra-die and five interface metal layers; (b) shows HB geometry. TOP cell views map local M1--M6 to M12--M7 while retaining pin XY geometry. Layer order, preferred routing and PG directions, and instance orientation are specified separately.

\begin{figure*}[t]
\centering
\includegraphics[width=0.97\textwidth]{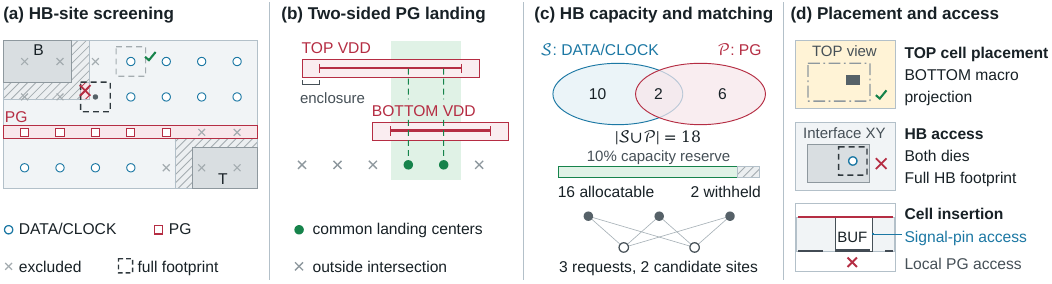}
\caption{Physical design constraints. (a) HB-site screening against BOTTOM (B), TOP (T), and PG. (b) Two-sided PG landing on same-supply metal. (c) HB capacity and matching: $\mathcal S/\mathcal P$ denote DATA/CLOCK- and PG-compatible sets. (d) Placement and access: cell placement, HB access, and cell insertion use different constraints.}
\Description{Four constraint examples show legal XY sites for each HB type, intersected PG landing intervals, eighteen distinct sites versus a locally deficient candidate graph, and the different domains around macros. Legal cross-die projection overlap need not permit HB access or a new buffer's VSS connection.}
\label{fig:resources}
\end{figure*}

DATA/CLOCK connects M6 to M7 through five local vias and one HB; PG connects B\_TM2 to T\_TM2 through two local vias and one HB. Both TM2 layers carry horizontal interface PG rails. Intra-die BOTTOM routing uses M2--M6 (three horizontal, two vertical layers); TOP uses M7--M11 (three vertical, two horizontal), with cell-pin access on M1 and M12, respectively. Vertical routing on BOTTOM and horizontal routing on TOP therefore have fewer intra-die routing layers, motivating direction-specific congestion analysis.

Rail widths and via enclosures set the required PG clearances. Final HB-site screening therefore uses the updated PG geometry. HB-interface timing arcs preserve driver-to-sink connectivity, and post-route RC extraction captures the realized interdie stack and the intra-die wires on both dies.

\subsection{HB-Site Screening}
Within the common die outline, grid $\mathcal H$ contains points $\phi+(kp_x,lp_y)$, where $\phi$ is its XY origin, $p_x,p_y$ are pitches, and $k,l$ are integer indices. At site $h$, HB type $\rho$ requires $F_\rho(h)\cap O_\rho=\emptyset$: its complete footprint $F_\rho(h)$ must avoid the applicable obstacle set $O_\rho$, including die boundaries, both dies' macro-access exclusions, and PG clearances (Figure~\ref{fig:resources}(a)). PG stripes can remove DATA/CLOCK eligibility while supplying power landings. For two-sided PG landing on supply net $n\in\{\mathrm{VDD},\mathrm{VSS}\}$,
\begin{equation}
 \mathcal P_n\subseteq\mathcal H\cap(L_B^n\ominus E_B)
                                  \cap(L_T^n\ominus E_T).
 \label{eq:landing}
\end{equation}
Here $\mathcal P_n$ is the set of legal PG HB sites for supply net $n$, and $B/T$ denote BOTTOM/TOP. The landing regions $L_B^n,L_T^n$ contain metal for the same supply on the two dies. The erosion operator $\ominus$ shrinks each region by its enclosure footprint ($E_B$ or $E_T$), leaving candidate HB centers (Figure~\ref{fig:resources}(b)). Clearance from the opposite supply and containment of the complete stack further restrict the intersection.

\subsection{HB Capacity and Matching}
Let $\mathcal S$ be DATA/CLOCK-compatible sites and $\mathcal P=\bigcup_n\mathcal P_n$. Physical capacity and capacity after reservation are
\begin{equation}
 C_{\rm phys}=|\mathcal S\cup\mathcal P|,\qquad
 C_{\rm alloc}=\lfloor(1-\gamma)C_{\rm phys}\rfloor+g.
 \label{eq:guard}
\end{equation}
Here $|\cdot|$ counts distinct sites, $0\le\gamma\le1$ is the reserved fraction, and integer $g$ counts released units of reserved capacity, with $0\le g\le C_{\rm phys}-\lfloor(1-\gamma)C_{\rm phys}\rfloor$. CU-3D deducts the reserve once, after obstacle filtering. To release capacity, it verifies that DATA and CLOCK HBs occupy distinct sites in $\mathcal S$ and that their total count plus the required PG minimum fits within $C_{\rm phys}$. PG requests must also match distinct eligible sites. The assignment enforces CLOCK-count bounds and the PG minimum against the same site pool. For a subset $U'$ of fixed one-HB requests, let $N(U')$ be their eligible-site union. Even with sufficient total capacity, $|N(U')|<|U'|$ prevents matching these requests to distinct sites (Figure~\ref{fig:resources}(c)).

\section{Partitioning and Placement}
\label{sec:placement}

\begin{figure*}[t]
\centering
\includegraphics[width=0.97\textwidth]{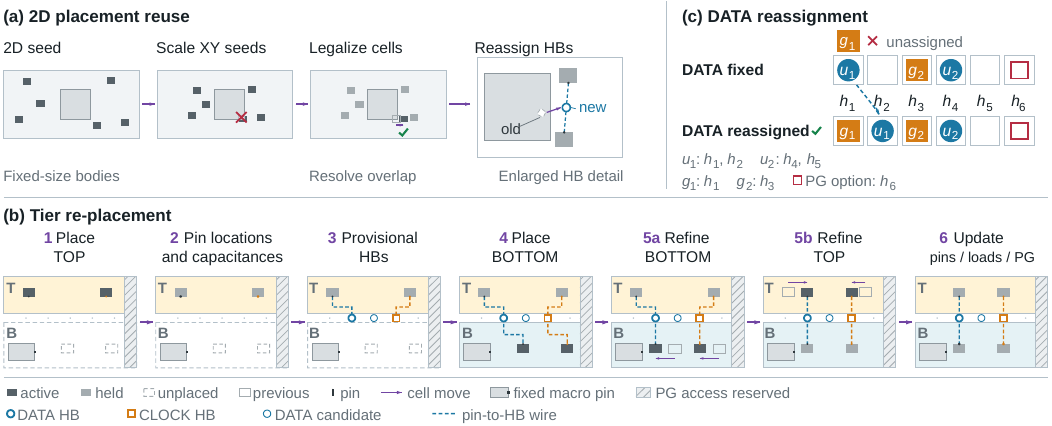}
\caption{Placement and DATA reassignment. (a) 2D placement reuse changes XY seeds, not cell dimensions. (b) Tier re-placement: place TOP, read pins/loads, assign provisional HBs, place BOTTOM, refine tiers (5a/5b), update pins/loads/PG (6); T/B: TOP/BOTTOM. (c) DATA reassignment moves $u_1$ to admit CLOCK group $g_1$, retaining $u_2$, $g_2$, and a PG option; $h_i$: site $i$.}
\Description{2D placement reuse retains fixed-size bodies; its last frame enlarges the HB and pin detail. Tier re-placement tracks the same four movable standard cells, fixed macro, and PG-access reservation. Previous-position outlines are not extra cells; dashed colored lines are provisional pin-to-HB wires. In the assignment example, h1 through h6 are sites, u1 and u2 are DATA requests, and g1 and g2 are CLOCK groups.}
\label{fig:routes}
\end{figure*}

\subsection{Macro Partitioning and Placement}
CU-3D automatically partitions and places macros: connectivity and area constraints guide tier assignment, followed by bounded, legal shifts of each tier's macro group. These rigid shifts increase cross-tier overlap of macro obstructions and HB clearances in the common XY plane, reducing the combined area unavailable to HBs. Macro dimensions, orientations, and internal pin, obstruction, and supply-rail geometry remain unchanged.

Available standard-cell area excludes same-tier blockages; HB sites must avoid both tiers' obstructions. Figure~\ref{fig:resources}(d) distinguishes \emph{placement and access}: a BOTTOM macro can lie below a legal TOP cell yet block an HB. Cell insertion additionally requires VDD/VSS and signal-pin access.

Placement refinement accounts for connections between macro pins and surrounding standard cells on the complete netlist. In a separate local refinement study using pin capacitance and estimated wirelength, matched-pin distances decreased but pre-CTS WNS worsened by 12.4 ps on BOTTOM and 27.9 ps on TOP. Shorter connections alone do not ensure better timing.

\begin{figure*}[t]
\centering
\includegraphics[width=.918\textwidth]{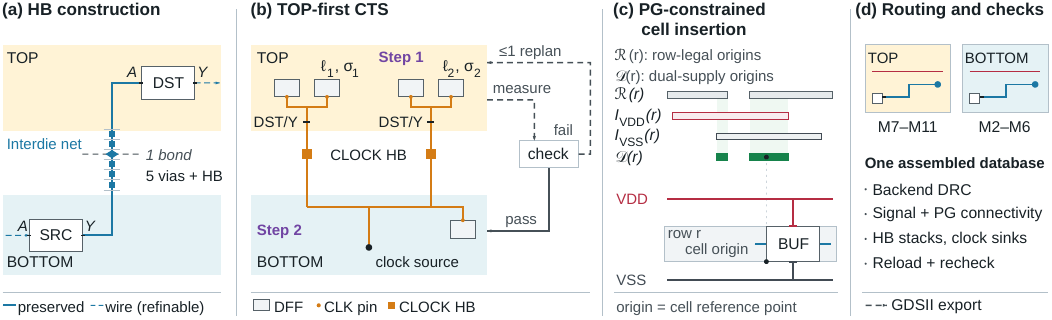}
\caption{CTS, PG, and routing. (a) HB construction: interdie net SRC/Y--DST/A. (b) TOP-first CTS: synthesize TOP subtrees and check latency $\ell_g$/skew $\sigma_g$, then synthesize BOTTOM. (c) PG-constrained cell insertion: intersect row-legal origins $\mathcal R$ with supply-access origins $I_{\rm VDD},I_{\rm VSS}$ to obtain $\mathcal D$. (d) Routing and checks in one database. Dashed: GDSII export with complete stream and cell views.}
\Description{Four modules show preserved interface stacks and refinable pin-to-HB wires, TOP-first CTS with at most one failed-check replan, dual-supply origin intervals for a candidate cell, and tier routing on one database. In a, five square local vias and one diamond HB join seven conductor levels; their spacing is schematic. The intervals in c are cell-origin sets.}
\label{fig:realization}
\end{figure*}

\subsection{Tier Partitioning}
Partitioning uses local area bounds within the reference 2D placement bins for 2D placement reuse, and die-wide bounds for tier re-placement. For total standard-cell area $A_b$ in region $b$ and usable tier areas $C_{b,B},C_{b,T}$, the area assigned to BOTTOM, $A_{b,B}$, must satisfy
\begin{equation}
 \max(0,A_b-C_{b,T})\le A_{b,B}\le\min(A_b,C_{b,B}).
 \label{eq:capacity}
\end{equation}
Tier-wide bounds allow cells to move across their original placement bins, so local balance need not split connected logic clusters. Fixed macros, placement rows, and access exclusions still limit usable area. Legalization and PG-access checks enforce these constraints at individual cell locations.

The partitioner compares HB-count violations, area-limit violations, non-CLOCK crossing demand, local area imbalance, and weighted cost, in that priority order. Lower-priority criteria break ties in higher-priority criteria. The final cost is
\begin{equation}
 J=\sum_{e:n_e^{\rm HB}>0}
 (w_e+w_e^t-w_e^c+w_{\rm HB}n_e^{\rm HB}).
 \label{eq:partition}
\end{equation}
For each crossing net $e$, $w_e$ is the base cut weight, $w_e^t$ is the timing-criticality penalty mapped to the partitioning netlist, $w_e^c$ is the congestion discount, and $w_{\rm HB}$ is the per-HB weight. Its driver, direction, and net class determine the required HB count $n_e^{\rm HB}$. This priority order favors resolving HB and area-limit violations over reducing $J$.

\subsection{2D Placement Reuse}
In 2D placement reuse, CU-3D scales reference coordinates, not cell or macro dimensions (Figure~\ref{fig:routes}(a)). It legalizes cells in each tier's rows, then locally refines placement on the full two-die netlist. Updated pin locations guide provisional HB reassignment within the fixed partition. Tight local area bounds can split connected clusters, increasing DATA crossings and consuming CLOCK-compatible sites.

CU-3D reserves whitespace for the cell-area increase measured in the 2D CTS reference. DATA reassignment and candidate expansion resolve site conflicts while preserving capacity for PG HBs. Unlike 2D placement reuse, tier re-placement does not retain the reference 2D cell arrangement.

\subsection{Tier Re-placement}
Tier re-placement recomputes standard-cell locations within each assigned tier. The TOP-first schedule uses TOP pin locations and clock-sink loads to guide HB assignment and BOTTOM placement, with BOTTOM macros and I/O fixed (Figure~\ref{fig:routes}(b)).

Steps 1--2 place TOP with unplaced BOTTOM cells temporarily disconnected, preserving macros, I/O, and PG-access reservations. Restoring connectivity yields TOP pin locations/capacitances and fixed BOTTOM macro/I/O pins. Steps 3--4 jointly assign provisional DATA/CLOCK HBs from the initial site inventory, excluding macro keepouts and retaining DATA candidates and PG capacity, then place BOTTOM toward these sites with TOP fixed.

Steps 5--6 refine BOTTOM then TOP on the complete netlist, fixing the other die, macros, HB interfaces, and PG-access reservations. Updated pin locations, loads, routing estimates, and PG geometry guide final HB-site screening and assignment. Costs account for pin spans, timing criticality, and displacement; timing analysis evaluates both dies' wires after each HB move.
\section{Joint HB Assignment}
\label{sec:joint}

\subsection{Joint HB Assignment Formulation}
CU-3D jointly assigns DATA and CLOCK HBs to legal sites, allowing at most one HB per site and reserving capacity for subsequent PG HB assignment. A CLOCK candidate $a=(d,s,h,G_a)$ specifies a clock domain $d$, a source $s$, an HB site $h$, and a sink set $G_a$. CU-3D screens candidates for fanout, diameter, reach, latency, skew, and capacitive load, estimated for direct fanout as
\[
 \widehat C_a=\sum_{i\in G_a}C_i+c_w\sum_{i\in G_a}\|p_i-h\|_1.
\]
Here $C_i,p_i$ are sink $i$'s current capacitance and XY location; $h$ also denotes the site's XY location. The coefficient $c_w$ is wire capacitance per unit length, and $\|\cdot\|_1$ denotes Manhattan distance. Buffered-tree estimates include buffer input loads and branch lengths. Each DATA request $u$ needs one HB; its candidate set $H_u$ retains the baseline site if legal. Binary variables $x_a$ and $y_{u,h}$ select CLOCK candidate $a$ and assign DATA request $u$ to site $h$, respectively. Writing $h(a)$ for candidate $a$'s site and setting $y_{u,h}=0$ outside $H_u$, we require
\begin{align}
 \sum_{a:i\in G_a}x_a&=1 \quad\forall i,\label{eq:cover}\\
 \sum_{h\in H_u}y_{u,h}&=1 \quad\forall u,\label{eq:assign}\\
 \sum_{a:h(a)=h}x_a+\sum_u y_{u,h}&\le1 \quad\forall h.\label{eq:occupancy}
\end{align}
Equations~\eqref{eq:cover}--\eqref{eq:occupancy} cover each clock sink $i$ once, assign each DATA request $u$ once, and allow one occupant per physical site $h$. The exact mixed-integer linear programming (MILP) mode applies CLOCK-count and regional limits, then prioritizes deviation of the selected CLOCK-HB count from its configured target, CLOCK timing, timing-critical DATA, remaining DATA, and control/I/O costs.

\subsection{Joint HB Assignment Algorithm}
\label{sec:joint-solvers}
\minisection{DATA reassignment} With identical CLOCK choices and constraints, feasible sets $\mathcal F_{\rm fixed}$ (fixed DATA) and $\mathcal F_{\rm joint}$ (movable DATA) satisfy $\mathcal F_{\rm fixed}\subseteq\mathcal F_{\rm joint}$ when every legal DATA baseline is retained. Setting $y_{u,\mathrm{baseline}(u)}=1$, where $\mathrm{baseline}(u)$ is $u$'s retained site, recovers each fixed solution. Solver mode governs how the enlarged feasible set is searched.

Figure~\ref{fig:routes}(c) contrasts \emph{DATA fixed} and \emph{DATA reassigned}. Its CLOCK groups $g_1,g_2$ define candidate sink sets $G_a$. Moving DATA $u_1$ from $h_1$ to $h_2$ frees the only site for $g_1$; $g_2$ and $u_2$ retain $h_3$ and $h_4$. Site $h_6$ remains available for subsequent PG HB assignment. The requests and site pool are unchanged.

\minisection{Solver selection} Algorithm~\ref{alg:joint} jointly selects CLOCK candidates and DATA sites. By model size, CU-3D selects exact MILP with ordered objective priorities, search-limited MILP, or spatial grouping with sparse site matching. Search-limited MILP fixes the target CLOCK count and retains legality constraints. For fixed CLOCK groups, spatial matching traces alternating paths from unmatched requests to a subset $U'$ with $|N(U')|<|U'|$; it adds legal candidate sites and rematches. The solver distinguishes proven infeasibility from an incomplete search.

\subsection{PG HB Assignment}
Joint DATA/CLOCK assignment reserves capacity for PG HBs. PG HB assignment then matches supply requests to distinct unoccupied legal sites within regional and distance limits. Let $N_D,N_C,N_P$ count assigned DATA, CLOCK, and PG HBs, respectively, and let $N_{P,\min}$ be the required PG count. An accepted assignment must satisfy $N_D+N_C+N_P\le C_{\rm alloc}$ and $N_P\ge N_{P,\min}$. Each PG HB connects the same supply on both dies.

\begin{algorithm}[t]
\caption{Joint DATA/CLOCK and subsequent PG HB assignment}
\label{alg:joint}
\noindent\textbf{Input:} Legal sites, pin locations/capacitances, demand, limits
\begingroup
\setlength{\leftmargini}{5.08mm}
\begin{enumerate}
\setlength{\itemsep}{0pt}
\setlength{\parsep}{0pt}
\setlength{\parskip}{0pt}
\item Build CLOCK/DATA candidate graph; retain legal baselines
\item Select size-adaptive mode; solve sink coverage and site occupancy
\item In spatial matching, add legal candidates and rematch
\item If incomplete, report unmet constraints or a search limit
\item Enforce capacity/count limits; assign PG HBs to remaining sites
\item \textbf{return} Feasible assignment or failed condition
\end{enumerate}
\endgroup
\end{algorithm}

\begin{figure*}[t]
\centering
\includegraphics[width=.92\textwidth]{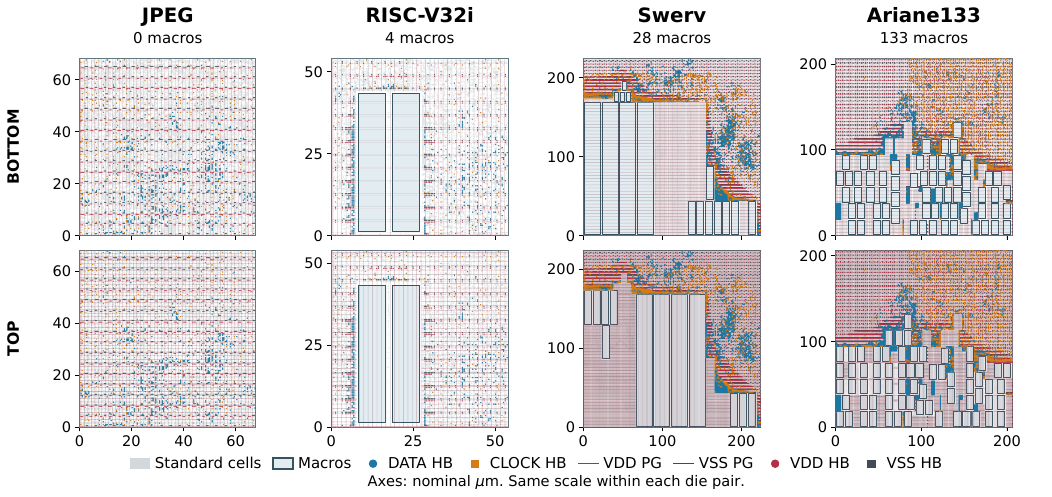}
\caption{Post-route layouts: HB centers and selected PG centerlines. BOTTOM/TOP scales match within each column. Our flow can guarantee the generated layout is DRC-clean and passes connectivity checks in the post-route implementation database.}
\Description{Eight layout panels compare both dies of JPEG, RISC-V32i, Swerv, and Ariane133. A single-row legend distinguishes standard cells, macros, DATA and CLOCK HB centers, VDD and VSS PG centerlines, and supply HB centers. The same interdie HB coordinates appear in both tier views; scales match within each design.}
\label{fig:layouts}
\end{figure*}
\section{CTS, PG, and Routing}
\label{sec:realization}

\begin{table*}[t]
\centering\fontsize{8}{9.2}\selectfont
\caption{2D/3D PPA and top-20 path composition. $P$: tool-reported power; $A$: per-die footprint; WL: wirelength.}
\label{tab:selected}\label{tab:overview_results}
\setlength{\tabcolsep}{4pt}
\begin{tabular*}{\textwidth}{@{\extracolsep{\fill}}llrrrrrrr@{}}
\toprule
Design & Placement & WNS (ns) & TNS (ns) & $P$ (mW) & $A$ ($\mu$m$^2$) & WL (mm) & Cell (\%) & Net (\%) \\
\midrule
Rocket & 2D & -0.428 & -321.400 & 42.413 & 5,971.3 & 204.56 & 64.0 & 36.0 \\
 & 2D reuse & -0.410 & -455.573 & 37.112 & 3,843.0 & 203.79 & 64.1 & 35.9 \\
\addlinespace[2pt]
AES & 2D & -0.003 & -0.005 & 6.911 & 1,672.5 & 47.31 & 86.3 & 13.7 \\
 & 2D reuse & 0.000 & 0.000 & 6.573 & 1,060.3 & 49.16 & 87.5 & 12.5 \\
\addlinespace[2pt]
Mini & 2D & -0.290 & -38.109 & 12.154 & 2,218.9 & 62.35 & 75.4 & 24.6 \\
 & 2D reuse & -0.268 & -51.340 & 10.294 & 1,404.5 & 55.50 & 76.9 & 23.1 \\
\addlinespace[2pt]
s35932 & 2D & 0.023 & 0.000 & 6.783 & 1,315.5 & 17.07 & 78.2 & 21.8 \\
 & Re-placement & 0.051 & 0.000 & 5.851 & 797.6 & 19.69 & 61.4 & 38.6 \\
\addlinespace[2pt]
Ariane133 & 2D & -2.877 & -10,388.600 & 233.445 & 71,074.4 & 1,632.51 & 38.1 & 61.9 \\
 & Re-placement & -2.876 & -18,240.000 & 215.131 & 42,640.6 & 1,489.83 & 37.1 & 62.9 \\
\bottomrule
\end{tabular*}
\par\smallskip\raggedright\fontsize{7.5}{8.5}\selectfont
Each row uses one implementation. Cell/net shares pool its 20 lowest-slack data paths; cell delay includes clock-to-output.
\end{table*}

\subsection{HB Construction}
HB construction preserves assigned locations, orientations, pins, and conductor stacks. Interface cells connect SRC/Y to DST/A through an interdie net; SRC/A and DST/Y connect upstream/downstream. PG stacks join same-supply landings. Routing may refine tier-local pin-to-HB wires while preserving HB geometry, connectivity, and TOP clock routes (Figure~\ref{fig:realization}(a)).

\begin{figure*}[t]
\centering
\includegraphics[width=\textwidth]{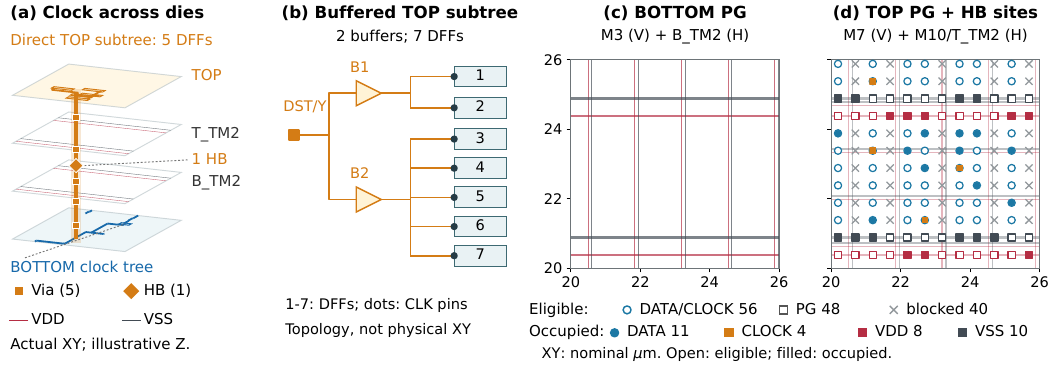}
\caption{Clock and PG structures in JPEG. (a,c,d) Measured 6-by-6-$\mu$m XY window. (a) Clock across dies: tier routes projected; Z schematic. Squares: five local vias; diamond: one HB. (b) Buffered TOP subtree: two parallel buffers serve seven DFFs, one buffer per path. (c,d) BOTTOM/TOP PG; (d) open/filled HB sites: eligible/occupied; crosses: excluded.}
\Description{Actual cross-die clock routes, PG, and HB sites share one spatial window. In a, five TOP cell bodies are DFFs; the BOTTOM body is the BOTTOM clock driver. The vertical stack contains three local vias below the single HB and two above. Via squares and the HB diamond follow PDK layer order, not physical Z spacing; within-tier routing layers are projected. Panel b is a separate TOP subtree with two parallel buffers and seven DFFs, one buffer per path.}
\label{fig:routed}
\end{figure*}

\subsection{TOP-First CTS}
The BOTTOM clock tree drives TOP subtrees through CLOCK HBs. CTS builds TOP first, with each subtree rooted at DST/Y. For group $g$, $\ell_g$ is the longest root-to-sink delay and $\sigma_g$ the longest-minus-shortest delay (Figure~\ref{fig:realization}(b)). These measurements, load, and geometry determine acceptance. Failed checks permit one replan: exclude the failed CLOCK-HB site, recompute the joint plan, and rebuild subtrees. BOTTOM CTS uses their delays on the full clock netlist, preserving TOP subtrees and HB interfaces.

For launch register $i$ and capture register $j$, let $\delta L_i,\delta L_j$ denote clock-latency changes. With other terms fixed, setup/hold slack changes are
\begin{equation}
 \Delta S^{\rm setup}_{i\to j}=\delta L_j-\delta L_i,
 \qquad \Delta S^{\rm hold}_{i\to j}=\delta L_i-\delta L_j.
 \label{eq:skew}
\end{equation}
In register chain $i\to j\to k$, delaying $j$ helps incoming setup but hurts outgoing setup; a shared branch affects every descendant. Physical sizing also changes upstream load and slew. Skew optimization must therefore evaluate all affected incoming and outgoing paths, including hold, in the complete clock graph rather than optimize each TOP subtree independently.

\subsection{PG-Constrained Cell Insertion}
Clock buffers and other standard cells added during timing optimization need legal VDD and VSS connections. For a candidate cell and orientation in row $r$, let $\mathcal R(r)$ be legal cell-origin positions and $I_{\rm VDD}(r),I_{\rm VSS}(r)$ the positions with legal local supply access. The insertion domain is $\mathcal D(r)=\mathcal R(r)\cap I_{\rm VDD}(r)\cap I_{\rm VSS}(r)$, with signal-pin access additionally required. Fixed placeholders and placement blockages reserve power-access space (Figure~\ref{fig:realization}(c)). Local pin access does not imply PG connectivity: a contacted rail may remain disconnected.

\subsection{Routing and Checks}
\label{sec:completion}
Routing obeys tier-specific layer ranges around assigned HB stacks and PG (Figure~\ref{fig:realization}(d)). Final checks evaluate backend DRC, signal/PG connectivity, HB-stack integrity, and clock-sink coverage in the assembled post-route database, followed by layout reload and recheck in a new Innovus session. Timing-coverage, setup/hold, and electrical-constraint reports remain separate. GDSII export uses complete stream and cell views.

\section{Experimental Evaluation}
\label{sec:experiments}

\begin{table*}[t]
\ifdefined\CUThreeDResultRowHeight\setlength{\extrarowheight}{\CUThreeDResultRowHeight}\fi
\def\CUThreeDPairedTables{1}
\begin{minipage}[t]{\columnwidth}
\ifdefined\CUThreeDPairedTables\else\begin{table}[t]\fi
\centering\small
\caption{JPEG variants. WNS/TNS: ns; WL: $10^3$ model $\mu$m; raw $P$: mW. All report DRC/connectivity 0/0.}
\label{tab:ablation}
\setlength{\tabcolsep}{2.5pt}
\begin{tabular*}{\columnwidth}{@{\extracolsep{\fill}}lrrrr@{}}
\toprule
Variant & WNS & TNS & WL & $P$ \\
\midrule
Bottom-first & -1.580 & -541.766 & 959.2 & 52.90 \\
Top-first, DATA only & -0.551 & -212.371 & 723.7 & 45.66 \\
Early joint HB assignment & -0.464 & -167.780 & 727.6 & 45.63 \\
+ Tier-wide area bounds & -0.118 & -54.046 & 573.9 & 41.81 \\
Timing-weighted & -0.175 & -139.920 & 606.1 & 41.73 \\
Fixed-sink refinement & -0.132 & -140.240 & 578.0 & 41.93 \\
Independent initializations & -0.219 & -203.370 & 628.8 & 43.83 \\
\bottomrule
\end{tabular*}
\ifdefined\CUThreeDPairedTables\else\end{table}\fi
\end{minipage}\hfill
\begin{minipage}[t]{\columnwidth}
\ifdefined\CUThreeDPairedTables\else\begin{table}[t]\fi
\centering\fontsize{8}{9.2}\selectfont
\caption{2D reuse (R) versus re-placement (P): non-CLOCK interface requests and tier-isolated horizontal/vertical (H/V) routing overflow.}
\label{tab:congestion}
\setlength{\tabcolsep}{2pt}
\begin{tabular*}{\columnwidth}{@{\extracolsep{\fill}}lrrr@{}}
\toprule
Design & Requests R/P & BOTTOM R $\to$ P & TOP R $\to$ P \\
\midrule
Rocket & 5,532/468 & 7/50 $\to$ 4/83 & 150/28 $\to$ 62/11 \\
JPEG & 6,230/578 & 0/1 $\to$ 0/6 & 47/5 $\to$ 20/0 \\
RISC-V32i & 2,257/354 & 95/62 $\to$ 0/0 & 19/2 $\to$ 3/0 \\
AES & 1,593/269 & 1/10 $\to$ 0/6 & 58/2 $\to$ 30/0 \\
Dynamic & 1,998/310 & 0/3 $\to$ 0/9 & 28/2 $\to$ 19/4 \\
Mini & 1,926/133 & 0/2 $\to$ 0/5 & 30/1 $\to$ 28/0 \\
\bottomrule
\end{tabular*}
\par\smallskip\raggedright
Same-stage global-route replay; entries give H/V overflow.
\ifdefined\CUThreeDPairedTables\else\end{table}\fi
\end{minipage}
\end{table*}

\subsection{Setup}
Ten designs cover logic and macro-rich processors: Rocket, JPEG, RISC-V32i, AES, Dynamic, Mini, s35932, gcd, Ariane133, and Swerv.

Synopsys Design Compiler and Cadence Innovus implement the designs on the ASAP7-based F2F model at TT, 0.70 V, 25~$^\circ$C. Nominal lengths and areas divide the $4\times$ model values by 4 and 16; nominal HB pitch is 0.5~$\mu$m. Both 2D and 3D flows use placement optimization, CCOpt, post-CTS setup optimization, timing-driven routing, and post-route setup optimization. Reported timing uses RC parasitics extracted from the routed layout.

We compare 2D placement reuse and tier re-placement. JPEG variants change tier \emph{placement} order and HB assignment, not TOP-first CTS. Backend DRC and connectivity checks use assembled post-route databases. Footprint is per die; power uses recorded activity. Wirelength sums routed segments: nominal mm for the 2D/3D PPA comparison, and $10^3$ model $\mu$m for JPEG variants (each unit equals 0.25 nominal mm). Layouts use final DEF coordinates and LEF dimensions; supply HB centers are checked against VDD/VSS nets.

\subsection{Post-Route Layouts and PPA}

All ten post-route designs have zero backend DRC violations and pass signal-connectivity checks; these checks include clock trees, PG, and HB stacks. Figure~\ref{fig:layouts} shows four representative layouts.

The layouts span 0--133 macros. Swerv has 3,544 DATA/CLOCK and 6,302 PG HBs (9,846 total), plus 56 macro PG terminals; Ariane133 has 4,379 DATA/CLOCK and 7,218 PG HBs (11,597 total). HBs are counted once across dies; macro PG terminals are not extra HBs. Both designs pass signal/PG connectivity. Swerv's tall macros narrow channels; Ariane's banks distribute access boundaries.

Table~\ref{tab:selected} shows five 2D/3D comparisons improving reported setup WNS, per-die footprint, and tool-reported power. With 2D reuse, Rocket/Mini gain 18/22 ps and AES reaches zero; re-placement gains 28 ps for s35932 and 1 ps for Ariane133 at the reported precision. Rocket/Mini improve WNS but worsen TNS across setup endpoints.

WNS measures the worst endpoint; TNS captures the aggregate setup deficit. Mini improves WNS with 11.0\% less routed wirelength. AES and s35932 also improve WNS, with 3.9\% and 15.4\% more wirelength, respectively. These aggregate results do not isolate which physical changes improve timing.

\subsection{Critical-Path Delay Analysis}
We independently select each 2D/3D report's 20 lowest-slack paths. Data delay comprises cell delay (including clock-to-output) and net delay, excluding clock latency and input arrival. Shares use pooled delays. Mini/AES reduce mean data delay by 56.2/19.7 ps; Ariane remains net-dominated in 2D/3D (61.9\%/62.9\% of data delay).

For matched comparisons, we select the connection with the lowest 2D slack present in both reports. Matches require identical start/end pins, transition polarities, path type, analysis view, and path group. Mini/AES cell delay falls 71.1/63.8 ps while net delay rises 9.2/13.5 ps. Internal logic may differ; these measurements do not isolate the cause.

The top-20 sets can contain different connections; matched comparisons fix endpoints and analysis conditions. Rocket's matched data delay falls 28.9 ps although its top-20 mean grows as input--register paths replace input--output paths. s35932's former worst connection moves to rank~82 in the 3D report.

\subsection{JPEG Placement and HB Assignment}
Bottom-first, top-first DATA-only, and early joint HB assignment reach $-1.580$, $-0.551$, and $-0.464$-ns WNS (Table~\ref{tab:ablation}). Joint assignment gains 87 ps despite 0.53\% more routed wirelength; total wirelength alone cannot explain timing.

Tier-wide area bounds reach $-0.118$ ns: WNS improves 346 ps, TNS 113.734 ns, wirelength 21.12\%, and raw power 8.38\%. DATA HBs (excluding control/I/O) fall from 6,211 to 561; CLOCK HBs rise from 293 to 471; PG is unchanged. Two 50-model-$\mu$m samplings reduce peak occupancy from 313/325 to 129/129. Fewer cut nets free sites without fixing the CLOCK-HB count.

Fixed-sink refinement and independent initialization lose 14 and 101 ps, respectively: early HB assignment needs subsequent cell-placement refinement. Further refinement reaches $-0.101$ ns; 18/20 worst low-HB paths stay within one die, shifting optimization toward cell delay, load, and clock latency.

\subsection{Congestion and Assignment Size}
Table~\ref{tab:congestion} reports tier-isolated routing replay. Re-placement reduces non-CLOCK requests 5.9--14.5$\times$ and TOP horizontal overflow in all six designs versus 2D reuse. Horizontal dominates all twelve TOP records; vertical dominates or ties in eleven BOTTOM records, consistent with layer counts and PG occupancy. H/V totals fall 53 to 26 for JPEG and 178 to 3 for RISC-V32i; Rocket BOTTOM vertical overflow rises 50 to 83.

\raggedbottom
\splitmaxdepth=0pt
\balance
The assignment contains 1,661 candidate variables for gcd, 3,043,683 for JPEG with 2D placement reuse, and 528,181 for JPEG with tier re-placement. They encode DATA request--site or CLOCK sink-group--site choices, not physical HBs. Sparse storage retains only candidates and connections. JPEG re-placement reduces model size 5.8$\times$ and routing demand.

\subsection{Clock and PG Structures}
\Cref{fig:routed}(a) shows clock routing across dies to five TOP DFFs.
\Cref{fig:routed}(b) shows a separate seven-DFF TOP subtree: two parallel buffers, one per path. HB assignment selects roots; TOP CTS builds branches.

\Cref{fig:routed}(c,d) show BOTTOM PG and TOP PG with HB sites in (a)'s XY window.
Of 144 sites, 56 support DATA/CLOCK, 48 PG, and 40 are excluded; 33 are occupied: 11 DATA, 4 CLOCK, 8 VDD, and 10 VSS HBs. Site eligibility differs from HB occupancy.

\section{Conclusion}
\label{sec:conclusion}
CU-3D carries HB-site and cell supply constraints through partitioning, placement, CTS, and routing. Constructed geometry determines HB capacity; pin locations and loads guide joint DATA/CLOCK assignment; CTS and routing preserve HB sites, stacks, and PG connections. All ten designs, including 28- and 133-macro processors, have zero backend DRC violations and pass signal-connectivity checks.
JPEG shows timing gains from DATA reassignment and tier-wide partitioning area bounds. In the Mini/AES matched-endpoint comparisons, cell-delay savings outweigh net-delay increases. Cell- and net-dominated paths motivate different optimization priorities. Site matching checks HB-assignment feasibility; post-route timing evaluates its benefit. Neither fewer bonds nor shorter total wirelength is a universal 3D objective.
Future work will feed matched-path cell, interconnect, and clock-phase costs into partitioning, HB selection, and macro placement refinement, and extend the physical design flow to multicorner timing, power integrity, and \mbox{thermal constraints}.\label{sec:bodyend}

\clearpage
\raggedbottom
\begin{acks}
This work is partially supported by
The Research Grants Council of Hong Kong SAR (No.~CUHK14211824 and T46-415/25-R).
\end{acks}

\balance

\end{document}